\documentclass[a4paper,fleqn,usenatbib,useAMS, twocolumn]{mnras}
\usepackage{ae,aecompl}
\usepackage{graphicx}	
\usepackage{amsmath}	
\usepackage{amssymb}	
\usepackage{amsbsy}
\usepackage{multicol}        
\usepackage{bm}		
\usepackage{listings}
\usepackage{alltt}
\usepackage{color}
\usepackage{booktabs}

\newcommand{\Msun}{\rm{M_{\rm \odot}}}

\newcommand{\Mbh}{\rm{M_{\rm BH}}}
\newcommand{\Mdm}{\rm{M_{\rm DM}}}

\title[Primordial black holes as dark matter]{Primordial Black Holes as Dark Matter: Constraints From Compact Ultra-Faint Dwarfs}

\author[Q. Zhu et al.]
{Qirong Zhu$^{1,2}$\thanks{E-mail: qxz125@psu.edu}, 
Eugene Vasiliev$^{3, 4}$, Yuexing Li$^{1, 5}$,  and Yipeng Jing$^{5}$
\vspace{0.5cm}\\
\parbox{\textwidth}
{\small $^{1}$Department of Astronomy \& Astrophysics; Institute for Cosmology and Gravity, 
The Pennsylvania State University, PA 16802, USA\\
$^{2}$Harvard-Smithsonian Center for Astrophysics, Harvard University, 60 Garden Street, Cambridge, MA 02138, USA\\
$^{3}$Rudolf Peierls Centre for Theoretical Physics, 1 Keble road, Oxford, UK, OX1 3NP\\
$^{4}$Lebedev Physical Institute, Leninsky prospekt 53, Moscow, Russia, 119991\\
$^{5}$Tsung-Dao Lee Institute; Department of Astronomy, Shanghai Jiao Tong University, 
800 Dongchuan Road, Shanghai 200240, China
}}

\begin{document}

\date{Accepted XXX. Received YYY; in original form ZZZ}

\pagerange{\pageref{firstpage}--\pageref{lastpage}} \pubyear{2017}

\maketitle

\label{firstpage}

\begin{abstract}

The ground-breaking detections of gravitational waves from black hole mergers by LIGO 
have rekindled interest in primordial black holes (PBHs) and the possibility of dark matter  
being composed of PBHs. It has been suggested that PBHs of tens of solar masses could 
serve as dark matter candidates. Recent analytical studies demonstrated that compact
ultra-faint dwarf galaxies can serve as a sensitive test for the PBH dark matter hypothesis, 
since stars in such a halo-dominated system would be heated by the more massive PBHs,   
their present-day distribution can provide strong constraints on PBH mass.
In this study, we further explore this scenario with more detailed calculations, 
using a combination of dynamical simulations and Bayesian inference methods. The joint 
evolution of stars and PBH dark matter is followed with a Fokker--Planck 
code {\sc phaseflow}. We run a large suite of such simulations for different dark matter 
parameters, then use a Markov Chain Monte Carlo approach to constrain the PBH 
properties with observations of ultra-faint galaxies. We find that two-body relaxation 
between the stars and PBH drives up the stellar core size,  
and increases the central stellar velocity dispersion. Using 
the observed half-light radius and velocity dispersion of stars in the compact ultra-faint 
dwarf galaxies as joint constraints, we infer that these dwarfs 
may have a cored dark matter halo with the central  density 
in the range of 1--2 $\rm{M_{\odot}/pc^3}$, and that the PBHs may have a mass 
range of 2--14 $\rm{M_{\odot}}$ if they constitute all or a substantial fraction of the dark matter. 

\end{abstract}


\section{Introduction} 
\label{sec:intro}

In the standard Lambda cold dark matter ($\Lambda$CDM) cosmology, the nature of the 
DM remains elusive. The possibility of DM being composed of primordial black holes 
(PBHs, \citealt{Hawking1971}), arising naturally from certain inflation models 
\citep[e.g.,][]{Frampton2010}, has been investigated in the context of galaxy formation 
early on \citep[e.g.,][]{Carr1974, Meszaros1975}, and it has not yet been completely 
ruled out \citep[see recent reviews by][]{Carr2016}. Recently, it was suggested that three 
mass windows,  around $5 \times 10^{-16}\Msun$, $2 \times 10^{-14}\Msun$ and 
25--100 $\Msun$, respectively, are still open for DM consists of PBHs \citep{Carr2017}.

The recent detections of gravitational waves by LIGO are believed to originate 
from merging black holes with masses of order 10--$50\,\Msun$, significantly 
higher than known BHs in Galactic X-ray binaries \citep{LIGO2016A, LIGO2016B, 
LIGO2017A, LIGO2017B}.  These discoveries have renewed theoretical interest 
in PBHs as a DM candidate. It was suggested by \cite{Bird2016} that PBHs with a 
mass around $30\, \Msun$ are loosely constrained by micro-lensing experiments 
and that the merger rate of binary PBHs is consistent with the LIGO detections. 

If true, PBHs in this mass range could be an attractive DM candidate to solve some 
well-known problems on the galactic scale \citep[see][]{Weinberg2015}, in particular 
the ``cusp vs. core problem'' (hereafter, we refer to dark matter made of primordial 
black holes as PBH-DM). The PBHs would produce DM cores instead of cusps in 
the central galaxy because any ``temperature inversion" present in a cuspy profile 
will lead to a removal the density cusp by increasing the velocity dispersion 
\citep{Quinlan1996}. Solutions to the ``cusp vs. core problem'' based on baryonic 
heating of DM are shown to be mass-dependent \citep{Pontzen2014, 
DiCintio2014, Chan2015}. The survival of a globular cluster in the ultra-faint dwarf
galaxy Eridanus II, however, hints to a DM core in a galaxy halo where any baryonic 
effect would be weak \citep{Amorisco2017, Contenta2017}. 

In addition, a cosmology model with PBHs as DM indicates a rather different evolution 
picture from that of other types of DM, such as \textit{warm dark matter}, 
\textit{fuzzy dark matter} \citep{Hu2000, Hui2017}, and \textit{self-interacting dark matter}
\citep{Vogelsberger2012}. Contrary to a common feature of suppressed small-scale 
perturbations in these models \citep{Vogelsberger2016}, fluctuations in the number 
density of PBH impose \textit{additional} small-scale power \citep{Afshordi2003}. 
As a result, low-mass PBH-DM halos will form ahead of  $\Lambda$CDM halos, 
while in \textit{warm/fuzzy dark matter}, the DM halo formation is considerably delayed, 
which has a strong consequence on the onset of reionization \citep{Yoshida2003, Hirano2017}. 

Despite these appealing features, it remains unclear what fraction of the DM can PBHs 
constitute, and what mass range can these PBHs have. Fortunately, a population of 
compact (with projected half-light radius of $\sim 30$ pc), ultra-faint ($\sim 1000\, L_{\odot}$) 
dwarf galaxies has recently been discovered \citep[e.g,][]{Koposov2015, Bechtol2015}, 
and they are expected to  provide stringent constraints on the PBH-DM. Using analytical 
approaches, \cite{Brandt2016} suggested that two-body relaxation between stars and 
PBH-DM would drive a substantial size evolution in galaxies, which is inconsistent with 
the observed small size of the ultra-faint dwarfs, while \cite{Koushiappas2017} argued 
that mass segregation would rearrange the stars into an otherwise detectable ring in 
the surface brightness profile of Segue I.

In order to improve these analytical studies, we combine two methods, Fokker--Planck 
(FP) simulations and Markov Chain Monte Carlo (MCMC) modeling, to study the evolution 
of galaxy halos containing both stars and PBH-DM and to compare with observations. 
We first use an FP code to perform a large set of galaxy simulations with different 
parameters for the PBH-DM to follow the interaction and evolution of the stars and PBHs, 
then we employ an MCMC method to constrain the PBH properties using  the observed 
half-light radius and velocity dispersion of stars in compact ultra-faint galaxies. Our modeling 
presents a significant improvement over the previous studies based on an analytical form 
of energy diffusion coefficient and other simplifying assumptions \citep{Brandt2016, Koushiappas2017}. 

This paper is organized as follows. In \S~\ref{sec:method}, we describe the numerical code 
and MCMC algorithm used in our simulations and analysis. In \S~\ref{sec:result}, we 
present the dynamical evolution of the stars and PBHs in the halo systems, and constraints 
from observations. We discuss the implications of our study and its limitations in 
\S~\ref{sec:discussion}, and summarize our main findings in \S~\ref{sec:conclusion}.

\section{Methods}
\label{sec:method}

\subsection{Numerical code for dynamical simulations}

In this study, we use an FP code, {\sc phaseflow}\footnote{{\sc phaseflow}
is part of the  AGAMA library for galaxy modeling,
available from \url{https://github.com/GalacticDynamics-Oxford/Agama/}.}
\citep{Vasiliev2017} to follow the dynamical evolution of stars and
PBH-DM in the galaxy halos.
{\sc phaseflow}  solves the one-dimensional FP equation
for the distribution function in energy, using a high-accuracy finite
element method. 
The code is well tested with problems such as core collapse and the formation of
Bahcall-Wolf cusp around a central massive black hole. {\sc phaseflow}
can handle multiple mass components, which enables our calculations of
the evolution of a halo of stars and DM consisting of PBHs. Once the
initial conditions of the two components are specified, {\sc phaseflow} evolves
the distribution function of all components
while at the same time updating the gravitational potential.
System diagnosis such as density profile and velocity
dispersions are computed
automatically, which greatly simplifies our analysis.
Throughout the paper, we set the Coulomb logarithm \citep{Binney2008}
$\ln \Lambda$ to be 15.

The stars in the DM halo are modeled with a Plummer sphere \citep{Plummer1911}:
\begin{equation}\\
\rho(r) = \frac{3M_{*}}{4\pi R_{0,*}^3} \bigg(1+\frac{r^2}{R_{0,*}^2} \bigg)^{-5/2},
\end{equation}
where the stellar system is characterized by the total stellar mass $M_{*}$ and its scale radius  
$R_{0,*}$, respectively. In this calculation,  we set $M_{*} = 10^3\, \Msun$, and  each star 
particle has an equal mass of $1\, \Msun$. 

The extended PBH-DM halo follows a Dehnen sphere \citep{Dehnen1993}:
\begin{equation}\\
\rho(r) = \frac{(3-\gamma)\Mdm}{4\pi R_{\rm 0, DM}^3} \bigg(\frac{r}{R_{\rm 0, DM}}\bigg)^{-\gamma} \bigg(1 + \frac{r}{R_{\rm 0, DM}}\bigg)^{\gamma-4}, 
\end{equation}
where $\Mdm$ is the total mass and $R_{\rm 0, DM}$ the scale radius of the halo, respectively. 
We choose the Dehnen sphere since it has finite mass in contrast to an 
Navarro-Frenk-White (NFW) profile 
\citep{Navarro1997}, while at the same time offers more flexibility than a Hernquist 
distribution \citep{Hernquist1990} or a truncated NFW profile. Note that the Dehnen sphere is 
a special case of the more general $\alpha$--$\beta$--$\gamma$ density profile 
\citep{Zhao1996}, where the density distribution is determined by the inner slope $\gamma$ and 
the outer slope $\beta$ and the steepness parameter $\alpha$,
which control how quickly the inner slope transits into the outer slope.

\subsection{Markov Chain Monte Carlo for parameter constraints}

\subsubsection{Observational data and the likelihood function}

The MCMC is an effective method to explore multiple key parameters in a complex system. In 
our model, we use both the stellar half-light radius and the stellar velocity dispersion of the 
compact ultra-faint dwarfs to constrain the parameter space, which requires $\sim$$10^5$ 
MCMC realizations. Fortunately,  {\sc phaseflow} is an extremely efficient code, and it takes 
only several seconds to run a single model. Thus, we combine a large grid of FP simulations 
performed using {\sc phaseflow} and MCMC calculations using the {\sc emcee} code. 
The {\sc emcee} code uses an affine-invariant ensemble sampler to efficiently 
random walk the parameter space \citep{ForemanMackey2013}. 

The likelihood function of our modeling is described in the following form:
\begin{equation}\\
\mathcal{L}  = \prod_{i=1}^{N} \mathcal{N} (r_{\rm h, M}, r_{\rm h, O}, \epsilon_{\rm h, O})\, \mathcal{N} (\sigma_{\rm M}, \sigma_{\rm O}, \epsilon_{\rm \sigma, O}), 
\end{equation}
where $\mathcal{N}$ is a normal distribution, $N$ is the sample size, 
$r_{\rm h, M}$ is the model output of 3D
half-mass radius, $r_{\rm h, O}$ and $\sigma_{\rm O}$ are the observed half-light radius
and velocity dispersion, respectively. The associated uncertainties are 
$\epsilon_{\rm h, O}$ and $\epsilon_{\rm \sigma, O}$. 

We use the top five ultra-faint dwarf galaxies compiled by \cite{Brandt2016} which have 
measurements of both size and stellar velocity dispersion available. 
The data is repeated here in Table~\ref{table:observation} for convenience but we refer 
the readers to \cite{Brandt2016} for details and references of each measurement. 
Note that there are discrepancies in the reported half-mass radius of the two newly 
discovered Ret II and Hor I between \cite{Koposov2015} and \cite{Bechtol2015}. 
In each case, we have chosen the smaller value of the two measurements. The reason 
for such a sample is based on the implicit assumption (inherited from \cite{Brandt2016}) 
that these five galaxies form a distinct class of faint dwarfs with similar size and 
velocity dispersion, which can be compared to a \textit{single} theoretical model. 
Lastly, we apply a factor of 1.3 to de-project the observed half-mass radius to 3D 
following \cite{Wolf2010}.

\begin{table}
\caption{A sample of observed compact ultra-faint dwarf galaxies used in our modeling from {\protect \cite{Brandt2016}}.}
\begin{center}
\begin{tabular}{c|c|c|c}
\toprule
Galaxy   Name     & Projected $r_{h}$ [pc]  & $\sigma_{*}$   [km/s]  & $L_{\rm V}\, [L_{\odot}]$ \\\hline
Wil  I                   & $25\pm6$    & $4.3^{+2.3}_{-1.3}$ & 1000 \\\hline
Seg I                   & $29^{+8}_{-5}$    & $3.9^{+0.8}_{-0.8}$ & 300 \\\hline
Seg II                   & $35\pm3$    & $3.4^{+2.5}_{-1.2}$ & 900 \\\hline
Ret  II                   & $32^{+2}_{-1}$    & $3.2^{+1.6}_{-0.5}$ & 1500\\\hline
Hor  I                   & $25^{+9}_{-4}$    & $4.9^{+2.8}_{-0.9}$ & 2000 \\
\bottomrule
\end{tabular}
\end{center}
\label{table:observation}
\end{table}

\subsubsection{Selections of prior distributions}

\underline{\bf {Total dark matter mass $\Mdm$}}\\

\noindent Compact ultra-faint dwarf galaxies from our sample have the least amount of stars
($\sim$$1000\, L_{\odot}$) ever known, but their total mass remains unknown. It was suggested 
that they may reside in the least massive halos where atomic cooling is efficient, and that
 their total mass is around $10^9\, \Msun$ \citep[][]{Sawala2015, Wheeler2015, Zhu2016}, 
 since such a halo mass is consistent with the $M_*$--$\Mdm$ relations at the low mass 
 end \citep{GarrisonKimmel2017}. Recently, \cite{Ma2017} resolved the formation of such 
 galaxies in their simulations, and reported that the least massive halo, which has a mass 
 of $10^8 \Msun$ at $z = 5$, had already stopped star formation since $z = 8$. On the 
 other hand, \cite{Read2016} modeled isolated dwarf galaxies and found that halos with 
 a total mass of $10^8 \Msun$ were able to match several dwarf satellites of the Milky Way.

Therefore, we choose a weakly informative prior of $\Mdm$, which follows a log-normal distribution:
\begin{equation}\\
\log(\Mdm) \sim \mathcal{N}(9, 0.5).
\end{equation} 

The choice of this prior is well justified in that the physics involved is relatively straightforward. 
These low-mass galaxies must be just massive enough, above the thresholds set by atomic 
cooling and UV background radiation, for stars to form \citep[e.g.,][]{Okamoto2008, Gnedin2014}. 
The width of the distribution also ensures that halos down to 2--$3\times 10^{8} \Msun$ are covered. \\

\noindent  \underline{\bf {DM scale radius $R_{\rm 0, DM}$}}\\

\noindent From the discussion below (\S~\ref{subsec:core_formation}), 
we argue that the inner density profile of the halo will form a core ($\gamma = 0$) due to relaxation
effect. However, the size of the scale radius needs to be treated as another free parameter in our 
model. This is largely due to a lack of self-consistent collisional N-body simulations from cosmological 
initial conditions. The assumption that the characteristic scales of the PBH-DM halos are similar to the 
$\Lambda$CDM models could lead to serious systematics. Thus, we adopt a log-uniform distribution 
of $R_{\rm *, DM}$ for the range from 100 pc to 2000 pc:
\begin{equation}\\
\log(R_{\rm *, DM}) \sim \mathcal{U}(\log(100), \log(2000)). \\
\end{equation} 

\noindent  \underline{\textbf{Stellar scale radius $R_{\rm 0, *}$ }}\\

\noindent The heating of stars due to two-body relaxation increases the scale radius
of stellar distribution roughly as $R_*\sim t^{0.5}$ \citep{Brandt2016},
which indicates that the present-day size does not 
depend sensitively on the initial value of $R_{\rm 0,*}$. We thus use a log-uniform distribution for 
$R_{\rm 0,*}$ for the range from 10 pc to 50 pc. 

\begin{equation}\\
\log(R_{\rm 0, *}) \sim \mathcal{U}(\log(10), \log(50)).
\end{equation} 

The lower limit corresponds to the largest Galactic globular clusters and the upper limit corresponds to the average 
size of the currently observed compact ultra-faint dwarfs. \\

\noindent  \underline{ \textbf{Primordial BH mass $\Mbh$}}\\

\noindent  Lastly, the mass of the PBHs is not well constrained \citep{Carr2017}.  
PBHs above $100\, \Msun$ can be ruled out by CMB observation and the Galactic 
wide binaries \citep[see][]{Carr2016}. However, since we are most interested in the range of 
[1, 100] $\Msun$, we assume  a log-uniformly distribution for $\Mbh$ in this range.

\section{Results}
\label{sec:result}

\subsection{Cusp to core transition in PBH-DM halos from relaxation effect}
\label{subsec:core_formation}

\begin{figure}
\begin{center}
\includegraphics[width=0.95\linewidth]{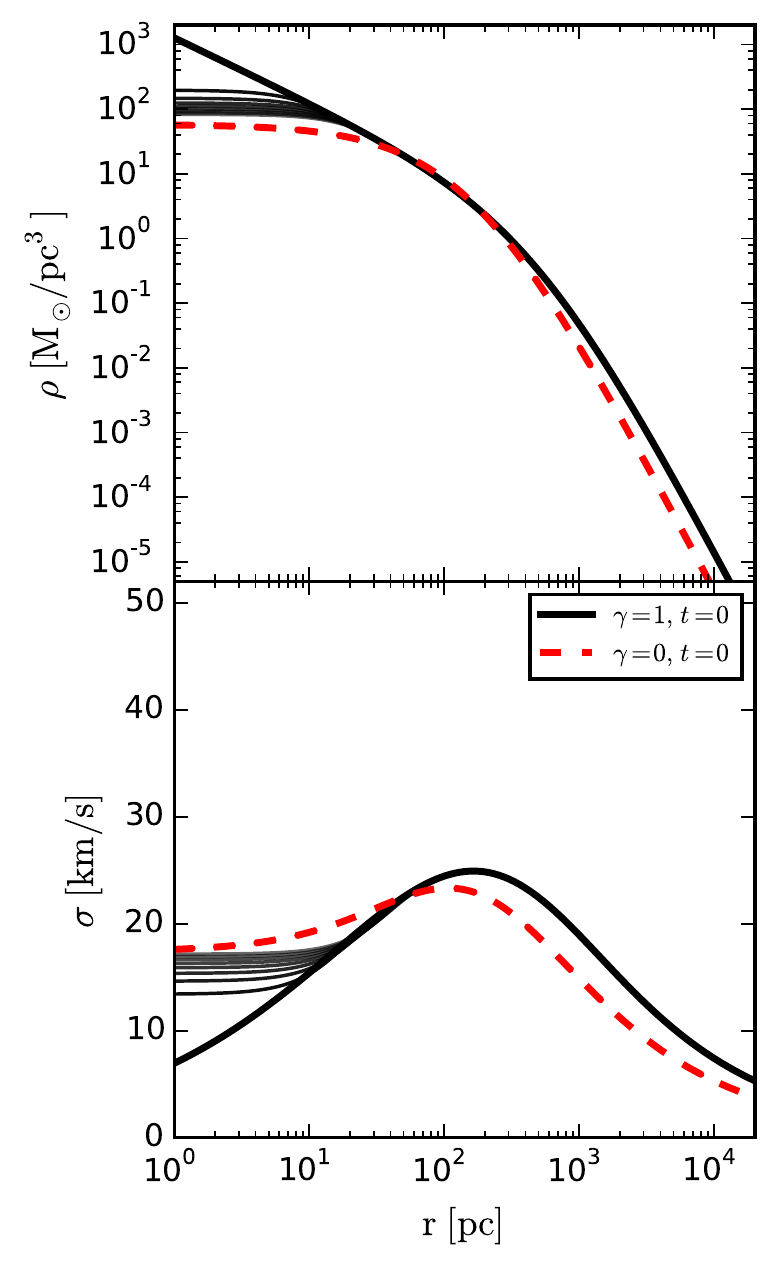}
\caption{The transition from cusp to core  (\textit{top panel}) and the increase of central velocity dispersion 
(\textit{bottom panel}) due to collisional relaxation effect in a PBH-DM halo. The DM halo has a mass of 
$2\times10^9\Msun$ consisting of $30\, \Msun$ PBHs. It has an initial density profile of $\gamma=1$ 
and $R_{\rm 0, DM}=500$ pc, and a central velocity dispersion of 8~km/s, as represented by the thick 
black curves in both panels. The thin lines show the evolution of the density profile and the projected 
velocity dispersion from $t=0$ to 12 Gyr with an interval of 1.5~Gyr. The red dashed lines shows the 
best matching PBH-DM halo of  $10^9\Msun$  with a core density profile of $\gamma=0$ and 
$R_{\rm 0, DM}=160$ pc, and an increased central velocity dispersion. }
\label{fig:galaxy_profile_evolution}
\end{center}
\end{figure}

In a CDM halo, the density profile usually shows a central density cusp with a slope in the range of 
$0 < \gamma < 2$, and the DM velocity dispersion peaks at the scale radius. However, if the DM is 
composed of PBHs,  two-body relaxation will soften
the central cusp, and as a result of the collisional heating, the velocity dispersion in the central region 
will increase while the density will drop at the same time. This process leads to the rapid formation of a 
core \citep{Quinlan1996} as a result of ``temperature inversion", in which the colder dense cusp will be 
heated by a hotter envelope.

Figure~\ref{fig:galaxy_profile_evolution} illustrates how this relaxation effect 
transforms a PBH-DM halo from a $\gamma=1$ cusp to a $\gamma=0$ core. The DM 
halo has a mass of $2\times10^9\Msun$ composed of $30\, \Msun$ PBHs, and an initial cuspy density 
profile of $\gamma=1$ and $R_{\rm 0, DM}=500$ pc. Collisional relaxation quickly removes the 
central density cusp and increases the central velocity dispersion. After 12~Gyrs, this PBH-DM halo 
can be approximated by a less massive ($10^9M_{\sun}$) one with a core density profile of $
\gamma=0$ and a characteristic scale radius of $R_{\rm 0, DM}=160$ pc, as shown by the red 
dashed curves.

In order to determine the time scale of the transition from cusp to core profiles, we set up cuspy DM halos
consisting of $30\, \Msun$ PBHs with a total mass in the range from $10^5$ to  $10^9\, \Msun$ at 
redshift $z=10$,  based on the mass-concentration relation from \cite{Diemer2015}. We evolve the 
PBH-DM halos using the FP code {\sc phaseflow}. These simulations show that the collisional relaxation 
effect removes the central cusp almost instantaneously, and the DM core grows quickly in size. For
instance, it takes only $\sim 0.05$~Gyr for all the halos to develop a core of $\sim 10$~pc, and 
$\sim 0.18$~Gyr to reach $\sim 20$~pc. If formed via hierarchical assembly, it would  be difficult 
for DM halos of $10^8 - 10^9\, \Msun$ to retain density cusps in the first place since the central 
density of their less massive progenitors has already been lowered. 
We speculate that the density profile of a PBH-DM halo, similar to self-interacting DM 
\citep{Vogelsberger2012}, could be described by a truncated singular sphere 
with a sizable core \citep{Shapiro1999}.  In the absence of a self-consistent treatment of 
density profile of PBH-DM halos, we thus use a cored  density profile with $\gamma=0$ 
for any initial DM scale radius $R_{\rm 0, DM}$ as our default choice for simulations in the 
following sections. In the Discussion Section, we will include a test using $\gamma=1$ to 
examine the outcome with cuspy density profiles. 

\subsection{Stars in PBH-DM halos}

\begin{figure}
\begin{center}
\includegraphics[width=0.95\linewidth]{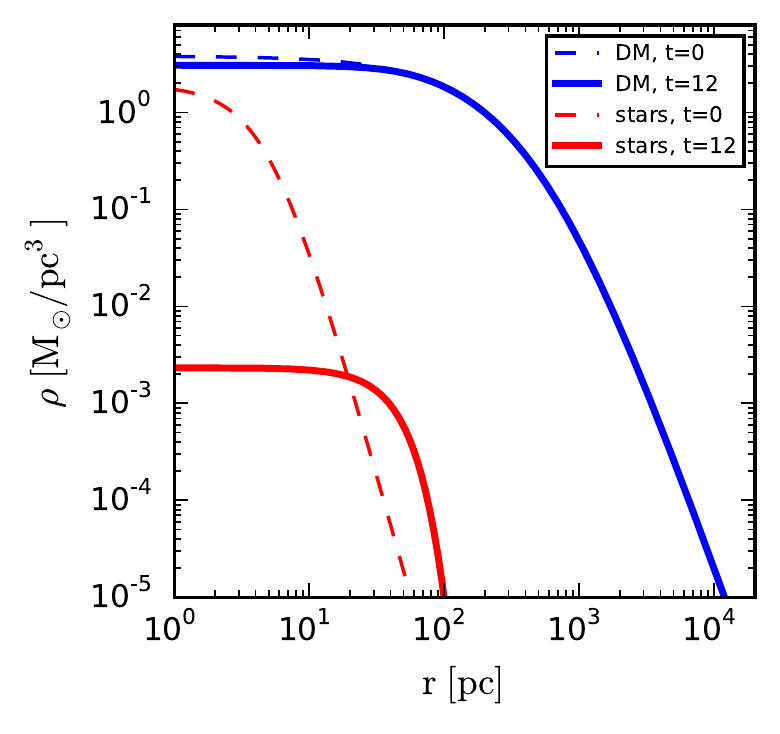}
\includegraphics[width=0.95\linewidth]{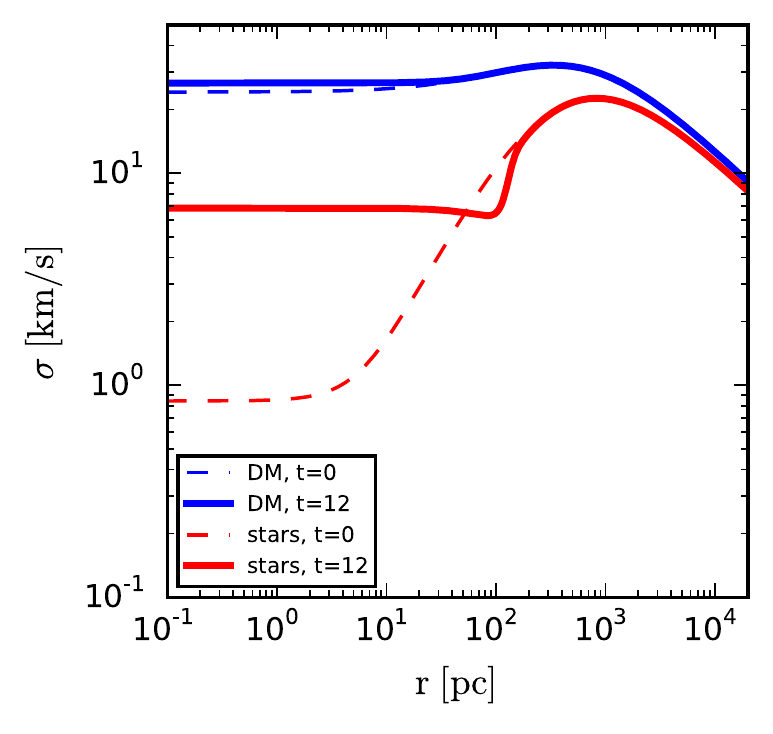}
\caption{Effects of two-body relaxation and energy exchange on the density profile (\textit{top panel}) 
and velocity dispersion (\textit{bottom panel}) of a two-component halo system with stars and PBH-DM. 
The halo has a total DM mass of $2\times10^{9}\, \Msun$ composed of 30 $\Msun$ PBHs, and a total 
stellar mass of $10^{3}\, \Msun$ consisting of $1\, \Msun$ stars. The blue curves represent the PBH-DM 
component at $t = 0$ (dashed lines) and $t = 12$~Gyr (solid lines), while the red curves represent the 
stellar components at $t = 0$ (dashed lines) and $t = 12$~Gyr (solid lines). }
 \label{fig:galaxy_profile}
 \end{center}
\end{figure}

\begin{figure}
\begin{center}
\includegraphics[width=0.95\linewidth]{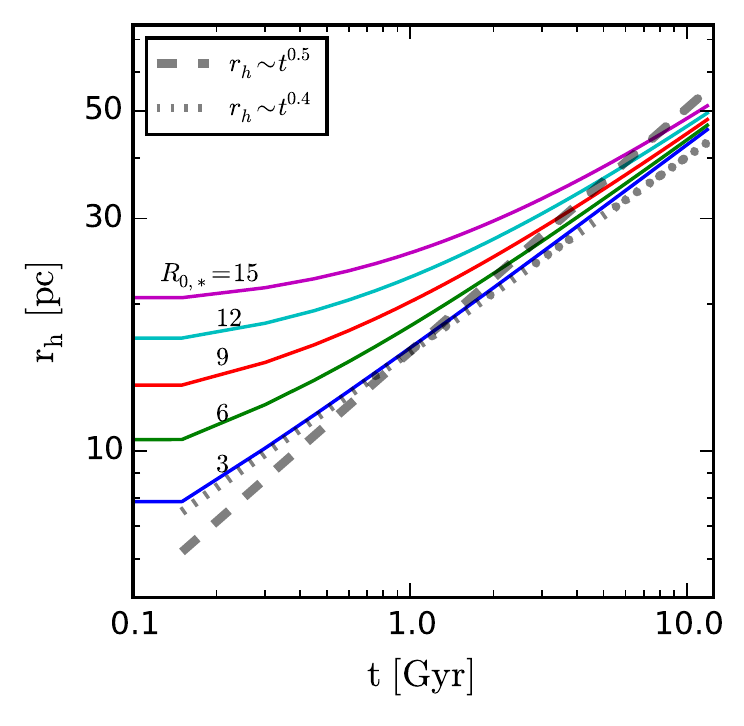}
\caption{The evolution of 3D half-light radius of stellar components with various initial conditions in a 
PBH-DM halo from our simulations, in comparison with analytical result from {\protect \cite{Brandt2016}}. 
The different color represents different initial condition with $R_{0,*}$ varying from 3, 6, 9, 12 and 15 pc, 
respectively. Our results show a $r_h \sim t^{0.4}$ size growth rate, slower than  $r_h \sim t^{0.5}$ 
reported by {\protect \cite{Brandt2016}}.}
\label{fig:halfm_radius_growth}
\end{center}
\end{figure}

In a system with two components of different masses, two-body encounters
and energy exchange would lead to change in the distribution  of density and energy. 
To investigate the evolution of a DM halo with stars and PBHs, we set up a DM halo with a mass of 
$2\times10^{9}\, \Msun$, which contains $10^{3}\, \Msun$ of stars, each being $1\, \Msun$, and the 
DM is composed of $30\, \Msun$ PBHs. The Plummer sphere for the stars has a total mass of 
$10^{3}\Msun$ and a scale radius of 5~pc, while the Dehnen sphere for the DM has a total mass of 
$2\times10^{9}\Msun$ and a scale radius $R_{\rm 0, DM} =$ 500~pc. The system is then followed 
with the FP code {\sc phaseflow}, and the results are shown in Figure~\ref{fig:galaxy_profile}. Due to 
the huge difference in both mass and size, the two components have different initialization in both 
density profile and velocity dispersion, and strong two-body relaxation and energy exchange leads to 
significant change in the stellar density profile and velocity dispersion. Because the PBH is much more 
massive than the star, heating from PBHs on the stars is the dominant driver for the evolution of the 
stellar component. After 12~Gyrs, the central stellar density drops by nearly 3 orders of magnitude, as 
shown in the upper panel of Figure~\ref{fig:galaxy_profile} , while the central stellar velocity dispersion 
increases by a factor of $\sim 8$, as in the lower panel.  

As the density profile of the stars slowly diffuses out, the half-light radius increases. This process 
depends on the density and velocity dispersion profile of the PBH halo. 
We note that in Figure~\ref{fig:galaxy_profile}, the stellar velocity dispersion is well below that of 
PBH-DM at $t=12$ Gyr. Hence, equipartition of energy between the two mass species, i.e. 
$\sigma(m)^2 \propto m^{-1}$, has not been achieved. 
 
As first pointed out by \cite{Spitzer1969}, full energy equipartition is possible only if the mass fraction 
of heavy species is below some critical value. In our setup, the total mass ratio between PBHs and 
stars is well above the critical value $\sim$$0.001(=0.16(1/30)^{1.5})$. As a result, the central regime 
will be devoid of stars, leading to only a ``partial equipartition". Recently, \cite{Trenti2013} and 
\cite{Bianchini2016} reported partial equipartition in their direct N-body simulations of globular clusters.  
Of course, our example is quite extreme since the mass density profile is completely dominated by 
PBHs at almost all radii.

The size increase due to heating of PBH appears to be slightly slower than the analytical result obtained 
by \cite{Brandt2016}. To further investigate this, we performed a set of FP simulations with {\sc phaseflow}
for different initial stellar density distributions, by varying $R_{0, *}$ from 3 to 15 pc,  within a fixed Dehnen 
sphere with a total DM mass of $2\times 10^9 \Msun$ and $R_{\rm 0, DM} =$ 500 pc. The resulting size 
evolution of the different stellar components is shown in Figure~\ref{fig:halfm_radius_growth}, in comparison 
with that of  \cite{Brandt2016}. Our simulations show a slower growth rate,  $r_h \propto t^{0.4}$, than 
that of $r_h \propto t^{0.5}$ by \cite{Brandt2016}. The heating rate of a less concentrated stellar component 
is also slower than those of more concentrated ones. For example, the magenta line on top of 
Figure~\ref{fig:halfm_radius_growth} shows the size evolution of a stellar core from $R_{0, *} = 15$ pc, 
which only approaches the asymptotic $t^{0.4}$ trend by the end of the integration. Our results suggest 
that the final size of the stellar core only depends weakly on the initial size, and that after $\sim 10$~Gyrs, 
a two-component PBH-DM halo of $2\times 10^9 \Msun$ would produce a stellar core of $\sim 50$~pc 
regardless of its initial size.

From Figure~\ref{fig:galaxy_profile}, the stellar density profile after $t = 12$~Gyrs is very smooth, we do not see any 
"ring profile" predicted by \cite{Koushiappas2017}. The discrepancy may be due to the different methods used to 
track the evolution of the halo system. The  FP simulations we performed with {\sc phaseflow} can track the two-body  
relaxation and energy exchange between the different mass components accurately, while the analytical formula 
used by \cite{Koushiappas2017} may not be able to follow the evolution of the system dynamically.

\subsection{Parameter space of PBH-DM halos from FP simulations}

\begin{figure}
\begin{center}
\begin{tabular}{c}
\includegraphics[width=\linewidth]{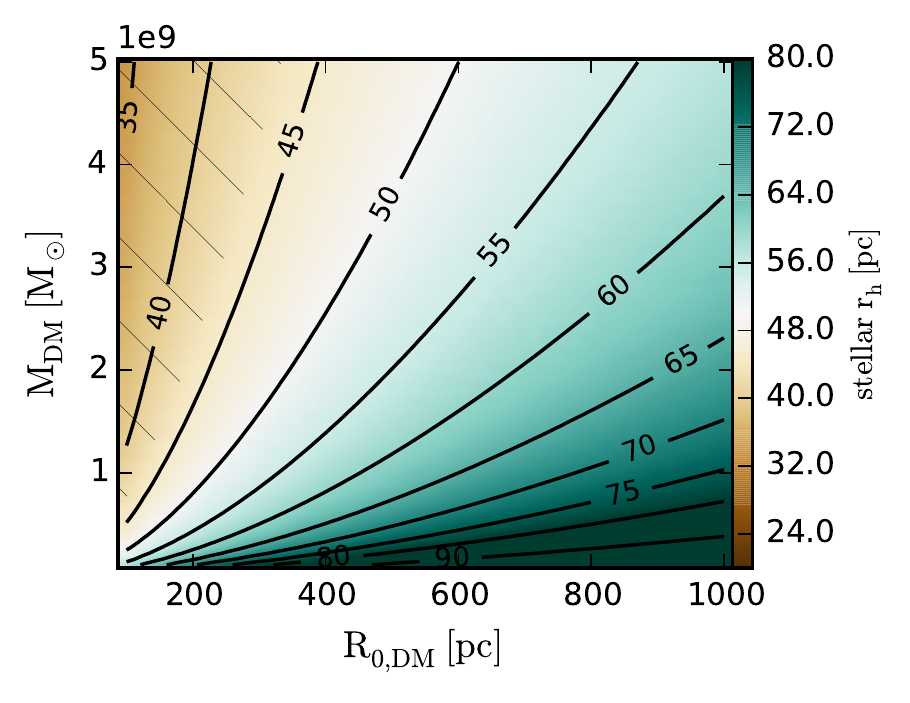}\\
\includegraphics[width=\linewidth]{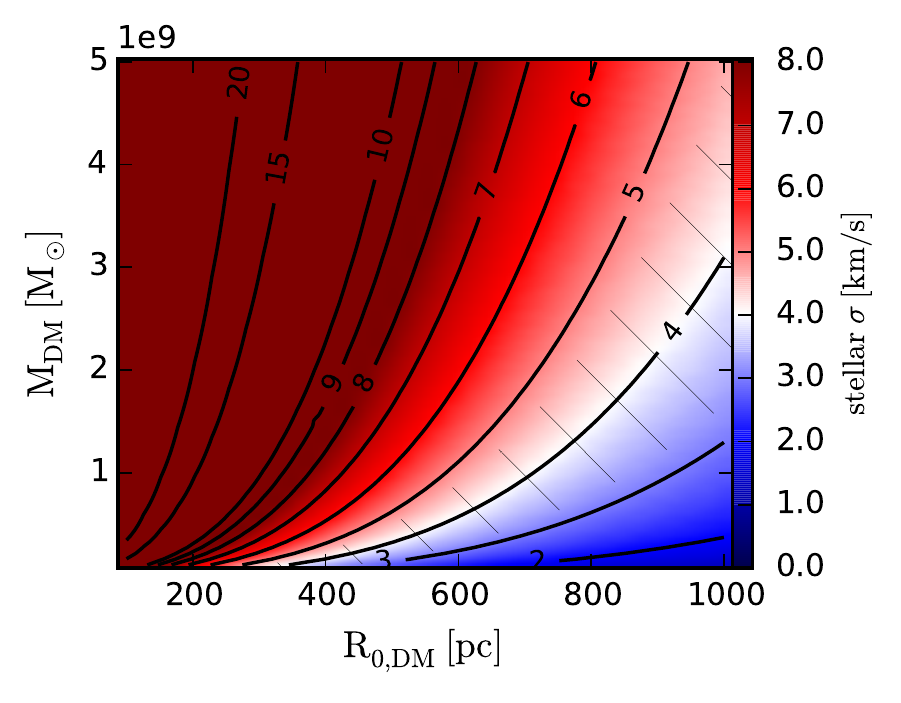}\\
\end{tabular}
\caption{The distribution of  3D half-mass radius (\textit{top panel}) and projected velocity dispersion 
 (\textit{bottom panel}) of the stellar components in the PBH-DM halo parameter space, from $2500$ 
 FP simulations using the {\sc phaseflow} code. There are 50 grids for each of the DM halo parameters,
  total DM mass $\Mdm$ and scale radius $R_{\rm 0, DM}$. The DM is assumed to consist of PBHs of
    $30\, \Msun$. The allowed regions by observations are highlighted and hatched in both panels which, 
    however, have little overlap with each other. }
\label{fig:galaxy_size_final}
\end{center}
\end{figure}

In order to explore the parameter space of DM halos of the observed compact ultra-faint dwarfs in 
Table~\ref{table:observation}, we use the FP code {\sc phaseflow} to perform 
a grid of dynamical simulations of the two-component systems consisting of stars and PBH-DM with 
different halo mass and size. We vary the total DM mass from  $10^8$ to $5\times10^9\, \Msun$, and 
vary the scale radius of the DM Dehnen sphere from 100~pc to 1000~pc.
We consider 50 values for each of the two DM halo parameters,
$\Mdm$ and $R_{\rm 0, DM}$, resulting in a total of $2500$ simulations.  
We assume the DM is composed of PBHs of  $30\, \Msun$, a total stellar mass of $10^3\, \Msun$, 
and an initial stellar scale radius of $R_{0, *} = 20$ pc. Each system is integrated for a duration of 
12~Gyrs.  

The final half-mass radius and the projected central velocity dispersion of the stars are shown in 
Figure~\ref{fig:galaxy_size_final}, as functions of  the DM parameter space of total DM masses 
$\Mdm$ and scale radius $R_{\rm 0, DM}$. The shaded regions in both the top and bottom panels 
indicate the allowed range of DM parameters from the observations of the compact ultra-faint dwarfs. 
From the stellar half-mass radius distribution in the \textit{top panel}, given the same total DM mass, 
the spreading of the stellar component is more efficient when the scale radius is larger. While with the 
same scale radius, less massive halos show more efficient heating. When the halo is less dense 
(either due to larger scale radius or lower mass), the initial velocity dispersions of both stars and PBHs 
are also much lower. In this case, the relative increase of stellar velocity dispersion is larger than in a 
more dense halo. The preferred range of $r_{h} \sim25 - 40$~pc from observations demand a narrow 
range of $R_{\rm 0, DM} < 400$~pc and $\Mdm < 5\times 10^9\, \Msun$. On the other hand, from the 
stellar velocity dispersion in the \textit{bottom panel},  the preferred range of $\sigma \sim 3-5$~km/s 
dictates a range of $300 < R_{\rm 0, DM} < 1000$~pc and $\Mdm > 10^9\, \Msun$. 

Clearly, it is difficult to reconcile the two constraints simultaneously. A smaller stellar size would require 
a DM halo with smaller scale radius, thus reducing the heating from relaxation by increasing the velocity 
dispersion. Given PBH mass of $30\, \Msun$ and a typical size of $\sim$40 pc (de-projected), there is 
little room in the plane of $\Mdm$ and $r_{\rm 0, DM}$ to meet both the constraints.

\subsection{MCMC constraints on PBH-DM from compact ultra-faint dwarfs}

\begin{figure*}
\includegraphics[width=0.85\linewidth]{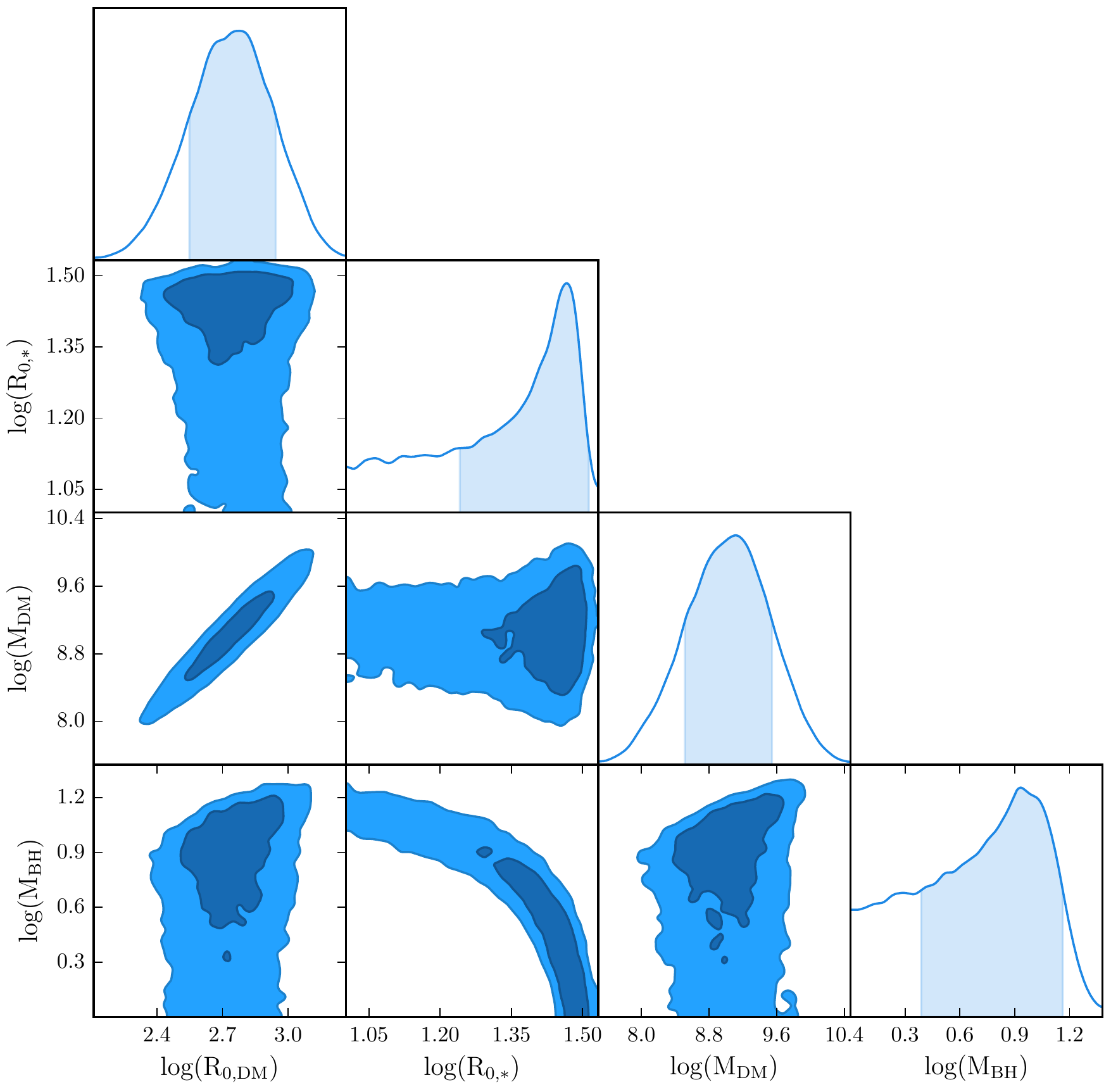}
\caption{Posterior probability distributions of the four PBH-DM parameters constrained by the stellar 
half-mass radius and stellar velocity dispersion of observed compact ultra-faint dwarfs: the DM scale 
radius $R_{\rm 0, DM}$, initial stellar component scale radius $R_{\rm 0,*}$,  total DM mass $\Mdm$, 
and PBH mass $\Mbh$. The contours indicate the 1-- and 2--$\sigma$ uncertainty region. This plot is 
generated using the Python package ChainConsumer \protect\citep{Hinton2016}. 
\label{fig:mcmc_result}}
\end{figure*}

In order to constrain the PBH-DM parameters from observations, we use the stellar half-mass radius 
and stellar velocity dispersion of the observed compact ultra-faint dwarfs in Table~
\ref{table:observation} as priors and perform a total of $10^5$ MCMC realizations using the {\sc 
emcee} code. The resulting posterior probability distributions of the four model parameters, the DM 
scale radius $R_{\rm 0, DM}$, initial stellar component scale radius $R_{\rm 0,*}$,  total DM mass $
\Mdm$, and PBH mass $\Mbh$, are shown in Figure~\ref{fig:mcmc_result}, respectively.  

The posterior surface in the plane of $R_{\rm 0, DM}$ and $M_{\rm 0, DM}$
can be understood with the help of Figure~\ref{fig:galaxy_size_final}, as a 
balance between the requirements of small sizes and small velocity dispersions.
The contours in the plane of $R_{\rm 0, DM}$ and $M_{\rm 0, DM}$ 
roughly covers a narrow region of constant DM density defined by 
$\Mdm \propto R^3_{\rm 0, DM}$. 

Moreover, the likelihood distribution of the initial stellar size $R_{0,*}$ and $\Mbh$ 
has two major features, which correspond to two major class of allowed configurations.
First, there is a vertical region with $\log(R_{0,*})$ close to the average size of the observed
dwarfs with $\log(\Mbh)$ close to 0. This is a class of models with little heating 
and, hence, insignificant size evolution for the stars. Second, there is a horizontal band where 
$\log(\Mbh)$ is close to 1 and $\log(R_{0,*})$ close to 1. This is the allowed 
parameter space where a substantial heating by PBHs expands the initially more
compact stellar distribution up to its present size.

From these distributions, the most probable results of the four parameters are listed in Table~\ref{tab:results}: 
$\rm {log (R_{\rm 0, DM}/pc}) = 2.77\, \Msun$, $\rm {log (R_{\rm 0,*}/pc}) = 1.47$ ,  $\rm {log (\Mdm/\Msun}) = 9.12$, 
and $\rm {log (\Mbh/\Msun}) = 0.93$. Note these results are similar to those obtained from simulations assuming 
cuspy initial DM density profiles, as shown in the table and discussed in \S~\ref{sec:discussion}.

Our results indicate that these galaxies must be DM-dominated: the posterior distribution of the
central DM density lies in a narrow range of 1--2 $\Msun/pc^3$ while the stellar density is
two orders of magnitude lower. We also experimented with models having no DM component at all. 
In this case, the velocity dispersion of stars would be much lower than the observed value for the
given combination of total stellar mass and half-mass radius. By contrast, globular clusters 
have a far larger stellar mass while occupying the same region 
in the $R_*$--$\sigma$ plane. The existence of DM also ensures the logic of using 
ultra-faint dwarf galaxies to constrain PBH mass is self-consistent.

The constraint on PBH mass in the massive end comes from a maximum heating rate, 
which is more stringent than that from the lower-mass end,   
mainly due to the assumptions upon which the stellar component  was initialized, 
as we discuss in \S~\ref{sec:discussion}. This can also be seen in the contour of 
$R_{*, 0}$ and $\log(\Mbh)$, which is a continuous distribution. Models 
with $R_{*, 0}$ between 20 and 30 pc are certainly allowed, with a modest heating
from PBH that brings the stellar size and velocity dispersion close to the observed
values. Note that the peak of $\Mbh$ does not correspond to the peak of
$R_{*, 0}$. Therefore, the likelihood surface in the plane of $R_{0,*}$ and $\Mbh$ 
is more informative, and the constraints on $\Mbh$ is limited by our 
prior knowledge of $R_{0,*}$.

The estimate can be improved if we have better prior information about the initial
size of the stellar component. There is a clear dichotomy in the spread in metallicity 
between dwarf galaxies and globular clusters of the same luminosities 
\citep[see][most of the observed globular clusters show $<0.1$ dex metallicity 
spread]{Willman2012}. The large metallicity spread ($>0.2$ dex) among the 
member stars in the dwarf galaxies could impose more stringent constraints 
on the initial stellar component size. Gas confined in a small volume would lead 
to efficient mixing due to multiple supernova events, hence a more uniform 
metallicity distribution is incompatible with observations. If the initial size $R_{0,*}$ is 
above 20 pc, then $\Mbh$ would fall below 10  $\Msun$.

\section{Discussions}
\label{sec:discussion}

\subsection{Model uncertainties}

\begin{figure*}
\includegraphics[width=0.85\linewidth]{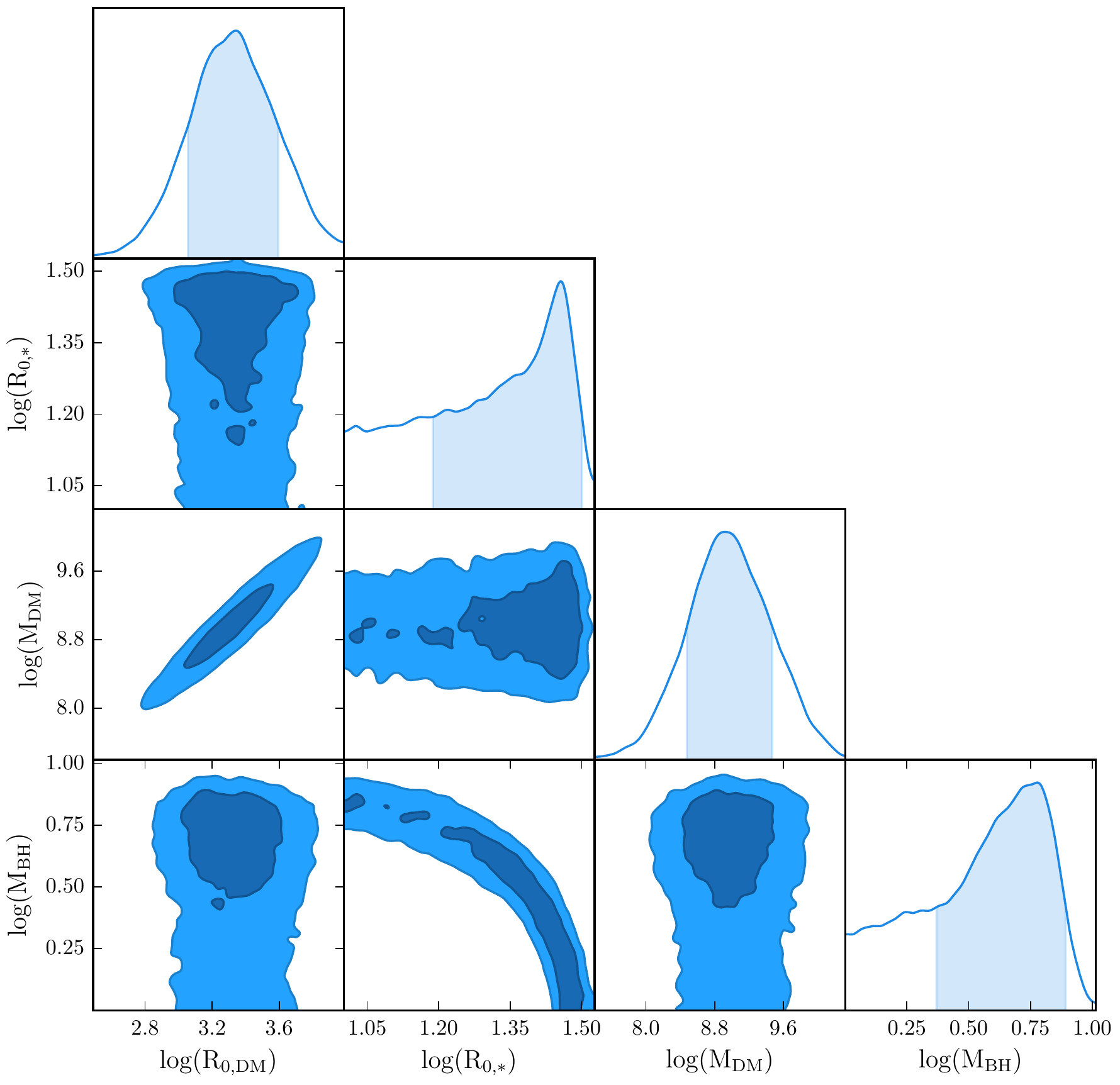}
\caption{Same as Figure~\ref{fig:mcmc_result}, but for a Dehnen halo with
a cusp ($\gamma=1$) at $t = 0$. The prior distribution of $\log(R_{\rm0, DM})$
is rescaled to $\sim \mathcal{U}(\log(200), \log(10000))$ in order to cover the allowed 
PBH halo parameter space. 
\label{fig:mcmc_result_cuspy_halo}}
\end{figure*}

In our models, the strongest prior is the total DM halo mass, which fulfills the requirement that these 
halos should be just slightly more massive than the minimum mass required for star formation. 
For the other three free parameters, we have adopted rather agnostic prior distributions. 
The initial distribution of stellar component, which follows a Plummer sphere, is a reasonable assumption.

Our model assumes that DM halo has a density core. This is a reasonable assumption given that the 
collisional effect on a cusped density profile is well understood. To demonstrate this, we have run 
another set of MCMC realizations with a density cusp ($\gamma=1$) at $t = 0$. As shown in 
Figure~\ref{fig:mcmc_result_cuspy_halo}, the final distributions of the four parameters are close to 
those in Figure~\ref{fig:mcmc_result}, although he peak of $\log(\Mbh)$ drops slightly from 0.93 to 0.78.
 This is due to extra heating  as the central velocity dispersion is low fora cuspy density profile. 
 A comparison of the marginalized posterior distributions for both cored and cuspy profiles is given 
 in Table~\ref{tab:results}.

\begin{table}
    \centering
    \caption{Results from $10^5$ MCMC realizations assuming cored and cuspy DM density profiles}
    \label{tab:results}
    \begin{tabular}{ccccc}
        \toprule
		model & $\rm{log(R_{0,DM})}$ & $\rm{log(R_{0,*})}$ & $\rm{log(M_{DM})}$ & $\rm{log(M_{BH})}$ \\ 
		\hline
		core & $2.77^{+0.17}_{-0.22}$ & $1.47^{+0.05}_{-0.23}$ & $9.12^{+0.42}_{-0.61}$ & $0.93^{+0.23}_{-0.54}$ \\ 
		cusp & $3.34^{+0.26}_{-0.28}$ & $1.46^{+0.05}_{-0.27}$ & $8.92^{+0.55}_{-0.44}$ & $0.78^{+0.11}_{-0.41}$ \\ 
		\bottomrule
    \end{tabular}
\end{table}

Our choice of prior distribution of total DM mass also a reasonable one. As illustrated in Figure~\ref{fig:mcmc_result}, 
the total DM halo mass $\Mdm$ does not have strong correlation with either $\Mbh$ or $R_{*, 0}$.  
The covariance between $\Mdm$ and  $R_{\rm 0, DM}$ covers a region 
defined by a constant DM density in the range of 1--2 $\Msun/pc^3$. In contrast, PBH mass is strongly 
correlated with the initial size of the stellar component. Therefore, our constraints on PBH mass is 
unaffected by the prior distribution of total DM mass. Moreover, if more compact ultra-faint dwarfs 
are to be discovered in the near future, their cumulative number could impose yet another constraint 
against the possibility of massive $\Msun$ PBHs as DM.  

There are several limitations in our approach need to be stressed. First, we treat the halos in isolation 
while in reality they are in highly tumultuous galactic environments, subjected to various mass loss 
processes \citep{Zhu2016}. Second, we use the measurements of the compact ultra-faint dwarfs as 
an input for the likelihood function and treat the reported uncertainties as the widths of Gaussians in
 the likelihood function. We have used a smaller half-mass size for Ret II and Hor I 
between those reported by \cite{Koposov2015} and \cite{Bechtol2015}. Accurate determination of 
the membership of stars in such faint galaxies is an extremely challenging task.  
In addition, unresolved binary stars may significantly inflate the observed velocity 
dispersion in these cold systems, although they are unlikely to completely dominate 
the measurements \citep{McConnachie2010}. Should the true size of these stellar 
systems change, our modeling and conclusions need to be revisited accordingly.

In our model, there is a built-in bias which puts more emphasis on the more massive PBH 
 for the following reason. If the collisional relaxation effect is negligible, only a limited region 
around $r_{h}=40$ pc and $\sigma=4\ {\rm km/s}$ is allowed. On the other hand, if the model 
starts with a smaller initial size (with a smaller velocity dispersion), a modest heating will 
bring both the size and the velocity dispersion closer to the observed values. Therefore, 
\textit{some} heating effect can be slightly more favored than \textit{no} heating effect if
there are more models with modest heating allowed than those with little heating.

\subsection{Comparison with previous works}

We believe that our analysis improves upon the previous calculations 
by \cite{Brandt2016} and \cite{Koushiappas2017} by using a FP code and an 
MCMC approach. In particular, the FP code self-consistently evolved the two 
components system with two mass species. In the latter study, the velocity 
dispersion profiles of the two components are assumed to be the same, 
while our modeling clearly shows this is not the case 
using the FP code. 

Moreover, our calculations show that the heating rate from 
the massive species on the light species is slightly less efficient than $t^{0.5}$, 
a result of partial equipartition of energy due to the dominant contribution from PBHs. 
In addition, the combination of the FP code and a MCMC approach is able to utilize both  
size and kinematics of the observed ultra-compact dwarf galaxies by efficiently 
sampling the parameter space.

We note that the stellar density profile by the end of the 
integration always contains a ``diffused" core in our FP calculation. 
This is not the same as the ring profile predicted by 
\cite{Koushiappas2017}. The main reason for this discrepancy is
that the size evolution formula used by \cite{Koushiappas2017} may not
be accurately applicable at all radii. On the other hand, the spreading out of
the surface density profiles are very similar to \cite{Koushiappas2017}.

Our model assumes that the DM is completely composed of PBHs. It was suggested by 
\cite{Brandt2016} that PBH above 10 $\Msun$ may be firmly excluded as the primary candidate 
of DM in dwarf galaxies.  The constraint from \cite{Koushiappas2017} is slightly stronger, 
with masses greater than 6 $\Msun$ excluded. The constraint derived from our analysis is slightly 
less restrictive compared to these two studies.

\subsection{Observational signatures of PBHs with millisecond time delay}

Our analysis is consistent with the suggestion that PBHs in the mass range of 25--100 $\Msun$ are 
ruled out as the main component of DM \citep{Brandt2016, Carr2017}. If the LIGO detections are 
from mergers of PBH binaries of $\sim 30\, \Msun$,  it is only possible if they are in the massive tail 
of an extended PBH mass function \citep{Magee2017}. However, any PBHs with a mass substantially 
higher than $10\, \Msun$ would violate the constraints from the dynamical analysis of dwarf galaxies.  
In the near future, aLIGO observations may soon provide constraints on PBHs around $10\, \Msun$.

It remains possible that \textit{some} fraction of the DM may be PBHs in this mass range. Apart from 
the contribution to the total DM budget, the existence of PBHs will  provide important constraints on 
certain inflation theories. In addition, gravitational lensing due to $\sim$10 $\Msun$ PBHs will induce 
a time delay on the order of milliseconds \citep{Mao1992, Munoz2016}. There are at least two known 
astrophysical phenomena that can produce detectable signals on this rapid time scale. 
One is fast radio burst \citep{Lorimer2007}. If fast radio bursts have indeed cosmological 
origin \citep{Spitler2016}, then one would expect repeated bursts with millisecond delay 
from the same location of the sky within a large number \citep{Fialkov2017} of bursts 
due to the PBH lenses \citep{Munoz2016}. 

Another promising source in the kHz regime for the 
detection of PBHs would be gravitational waves from stellar mass BH mergers. 
Similar to the mechanism of femtolensing of $\gamma$-ray burst \citep{Gould1992}, 
millisecond delays could induce a characteristic interference pattern in the detected 
waveforms in the merger phase due to the interference of two lensed images. 

Perhaps gravitational waves are better sources than radio bursts since they do not 
suffer from interstellar dispersions. This technique could be feasible with the next generation 
detectors such as the ``Cosmic Explorer" \citep{Abbott2017}. With a large number of 
gravitational wave detections, we could place a strong constraint on the mass 
fraction of PBHs in $\Omega_{\rm M}$,  since the lensing optical depth is at the same 
order as the mass fraction of PBHs for cosmological sources \citep{Press1973}.

\section{Summary}
\label{sec:conclusion}
 
In this study, we improve the recent analytical work of
\cite{Brandt2016} and \cite{Koushiappas2017}
on the investigation of the possibility of DM consisting entirely of
PBHs, by combining Fokker--Planck simulations with Markov Chain Monte
Carlo realizations. 
The former method, implemented in the code {\sc
phaeflow}, accurately follows the joint evolution of stars and PBH-DM
driven by two-body relaxation, for a given
choice of parameters. By combining it with the MCMC approach, we
explore the parameter space
and use the half-light radius and central stellar velocity dispersions
of observed compact ultra-faint dwarf galaxies as joint priors in the
MCMC sampling to constrain the  PBH-DM parameters. 
Our findings can be summarized as follows:

\begin{itemize}

\item For a DM halo in the mass range of  $10^6  - 10^9\, \Msun$, if the DM is composed of 
PBHs entirely, collisional relaxation quickly transforms a cusp to a cored density profile.

\vspace{0.5cm}

\item For a system with stars and PBH-DM, two-body relaxation between the stars and 
PBH-DM drives up the stellar core size,  and heating from the PBHs increases the central 
stellar velocity dispersion. The size evolution of the stellar core has a growth rate of 
$r_h \propto t^{0.4}$, which is slightly slower than that from the full energy equipartition 
($r_h \propto t^{0.5}$). 

\vspace{0.5cm}

\item The joint constraints from the observed half-light radius and central stellar velocity 
dispersions of compact ultra-faint dwarf galaxies leave a narrow parameter space of DM halos 
consist of PBHs. Results from the MCMC simulations suggest that the central density of DM halo 
has a narrow range of 1--2 $\Msun/pc^3$, and the mass of PBHs is most likely within 2--14 
$\Msun$, if the DM is completely made of PBH.

\end{itemize}

We note that our constraint of PBHs is slightly less restrictive than those of \cite{Brandt2016} 
and \cite{Koushiappas2017}. Although PBH with mass above 10 $\Msun$ can be excluded as 
the primary candidate of DM from our results, we emphasize that our models are inconclusive, 
and that the nature of DM remains elusive.

\section*{ACKNOWLEDGEMENTS}
It is our great pleasure to thank Avi Loeb, Lars Hernquist, Savvas Koushiappas and 
Peter M\'esz\'aros for stimulating discussions. YL acknowledges support from NSF 
grants AST-0965694, AST-1009867, AST-1412719, and MRI-1626251. The numerical 
computations and data analysis in this paper have been carried out on the CyberLAMP 
cluster supported by MRI-1626251, operated and maintained by the Institute for 
CyberScience at the Pennsylvania State University, as well as  the Odyssey cluster 
supported by the FAS Division of Science, Research Computing Group at Harvard 
University. The Institute for Gravitation and the Cosmos is supported by the Eberly 
College of Science and the Office of the Senior Vice President for Research at the 
Pennsylvania State University. EV acknowledges support from the European Research 
Council under the 7th Framework Programme  (grant 321067).


\bsp	
\label{lastpage}
\end{document}